\documentclass[11pt]{article}
\usepackage{graphicx}
\usepackage{amssymb}
\usepackage{epstopdf}
\usepackage{subfigure}
\usepackage{citesort}
\newcommand{\xvec}{ \vec{x} }
\title{Renormalization group approach to
the P versus NP question}
\author{S.N. Coppersmith, Department of Physics, University of Wisconsin--Madison,\\
1150 University Avenue, Madison, WI  53706}

\begin{document}
\maketitle

\begin{abstract}
This paper argues that the ideas underlying the renormalization group
technique used to characterize phase transitions in condensed
matter systems
could be useful for distinguishing computational
complexity classes.
The paper presents a renormalization group transformation
that maps an arbitrary Boolean function of $N$ Boolean
variables to one of $N-1$ variables.
When this transformation is applied repeatedly,
the behavior of the resulting sequence
of functions is different for
a generic Boolean function than for
Boolean functions that can be written
as a polynomial of degree $\xi$
with $\xi \ll N$ as well as for functions that depend on
composite variables such as the arithmetic sum of the inputs. 
Being able to demonstrate that functions are
non-generic is of interest because it suggests an avenue for
constructing an algorithm capable of demonstrating
that a given Boolean
function cannot be computed using resources that are bounded
by a polynomial of $N$.
\end{abstract}

\section{Introduction}
Computational complexity characterizes how the computational resources
to solve a problem depend
on the size of the problem specification~\cite{papadimitriou94}.
Two well-known complexity classes~\cite{complexityzoo}
are P, problems that can be solved
with resources that scale polynomially with the problem size, and NP,
the class of problems for which a solution can be verified with polynomial
resources. 
Whether or not P is equal to NP~\cite{cook71,levin72}
 is a great outstanding question in
computational complexity theory and in mathematics
generally~\cite{claywebsite,boppana90,sipser92,aaronson03,wigderson06}.

In this paper it is argued that a method known in statistical physics
as the renormalization group (RG)~\cite{kadanoff66,wilson71,wilson79,goldenfeld92}
may yield useful insight into the
P versus NP question.
This technique, originally formulated to provide insight into the
nature of phase transitions in statistical mechanical systems~\cite{kadanoff66,wilson71},
involves taking a problem with $N$ variables and then rewriting it as
a problem involving fewer variables.
Here, we will define a procedure by which a given
Boolean function of $N$ Boolean variables
is used to generate a Boolean function of $N-1$ variables, and
investigate the properties of the resulting sequence of functions
as this procedure is iterated~\cite{white92}.
The transformation used here is very simple --- the new function
is one if the original function changes its output value when a given
input variable's value is changed, and is zero if it does not.
It is shown that when this transformation is applied repeatedly,
the behavior of the resulting sequence of functions can be used
to distinguish
generic Boolean functions
from functions that are known to be computable using polynomially
bounded resources.

Any Boolean function $f(x_1,\ldots,x_N)$ of the $N$ Boolean variables $x_1,\ldots,x_N$
can be written as a polynomial in the $x_j$ using modulo-two addition.
This follows because the variables and function all can be only $0$ or $1$,
so $f(x_1,\ldots,x_N)$ can be written as
\begin{eqnarray}
f(x_1,\ldots,x_N) &=& A_{00\ldots 00}(1\oplus x_1)(1 \oplus x_2)
\ldots(1\oplus x_{N-1})(1\oplus x_N)\nonumber\\
&\oplus& A_{00\ldots 01}(1\oplus x_1)(1\oplus x_2)\ldots
(1\oplus x_{N-1})(x_N) \nonumber \\
&\ldots&\nonumber\\
&\oplus& 
A_{11\ldots 10}(x_1)(x_2)\ldots(x_{N-1})(1\oplus x_N)\nonumber\\
&\oplus& A_{11\ldots 11}(x_1)(x_2)\ldots(x_{N-1})(x_N)~,
\label{eq:general_form}
\end{eqnarray}
where $A_{x_1,\ldots,x_N}=f(x_1,\ldots,x_N)$.
As Shannon pointed out~\cite{shannon49},
the number of different possible functions is $2^{2^N}$
(this follows because each of the $2^N$ coefficients $A_{\alpha_1,\ldots,\alpha_N}$
can be either one or zero).
This is much larger
than the number of functions that can be computed using resources that
scale no faster than as a polynomial of $N$, which scales
asymptotically as $(CN)^t$, where $C$ is a constant and $t$
is a polynomial in $N$~\cite{riordan42,webcourse267}.
This counting argument demonstrates that almost all functions cannot
be evaluated using polynomially bounded resources
and hence are not in P.
However, it does not provide a means for determining whether or not
a given function can be computed with polynomial resources.

It is shown here that different classes of functions have different
behavior upon repeated application of a renormalization group transformation.
In analogy with well-known results
in statistical mechanics~\cite{goldenfeld92}, we interpret
functions exhibiting different behaviors after many
renormalizations as being in different phases.
Generic Boolean functions exhibit simple ``fixed point"
behavior upon renormalization,
and hence we claim that they comprise a phase.
A function that
can be written either as a low-order polynomial or as a function of
a composite variable such as the
arithmetic sum of the values of the inputs
yields non-generic behavior upon renormalization, and so is
in a non-generic phase.

We then discuss what would be needed to be able to use
the renormalization group approach to demonstrate
that a given Boolean function of $N$ variables cannot be evaluated with
resources that are bounded above by a polynomial in $N$.
This issue is relevant to the P versus NP question
because if we can identify a function in NP
that we can show is not in
P, then we will have shown that P and NP are not equal.
Some functions that are in P depend on the arithmetic
sum of the inputs, including 
MAJORITY, which is one if more than half the inputs
are nonzero and zero otherwise~\cite{razborov87}, and
DIVISIBILITY MOD $p$, which is one if the sum of the
inputs is divisible by an odd prime $p$~\cite{smolensky87,smolensky93},
and the renormalization group approach identifies
these functions as non-generic.
The renormalization group approach identifies low-order
polynomials as non-generic, and
some but not all low-order polynomials are in P.
Because there are functions in P that are the sum
of a low-order polynomial plus a small random component
that is nonzero on a small fraction of the inputs, and
because such functions will ``flow'' to the generic fixed
point upon renormalization,  P is not
a phase in the statistical mechanical sense.
Therefore,
there are functions known to be
in P that can be identified as non-generic only
because they are close
to a phase boundary in the sense that they differ from a low-order polynomial
on a small fraction of the inputs.
Thus, the renormalization group approach provides a means for understanding
why the P versus NP question is so difficult ---
showing that a function is not in P using
the renormalization group approach requires determining not only
that it is not in a non-generic phase but also that it
is not near a phase boundary, a task that appears
to require resources that grow
faster than
exponentially with $N$.
This superexponential scaling
means that the procedure proposed here cannot be used
to break
pseudorandom number generators, a difficulty that
would arise if the
procedure could be implemented with resources that scale
no faster than exponentially with $N$~\cite{razborov94natural}.
However, at this point we cannot prove that a given function is not
in P---our procedure distinguishes every function in
P of which we are aware from a generic Boolean function, but
we have not demonstrated that the procedure works for
all functions that are in P.

The paper is organized as follows.
Sec.~\ref{sec:RG} presents the transformation that maps a Boolean
function of $N$ variables into a Boolean function of $N-1$ variables.
Repeatedly applying this transformation yields a sequence
of functions, and in Sec.~\ref{sec:RG}
it is shown that (1)
if one starts with a generic random Boolean function, then the
resulting sequence of functions has the property that all functions
in it are nonzero for just about half the input configurations, (2)
applying the RG transformation $\xi$
times to a
function that is a polynomial of order less than $\xi$ yields zero,
and (3)
applying the RG transformation to functions that depend on
a composite variable such as the sum of the values of all the
inputs also yields a sequence of functions that differs from
from the result for a generic Boolean function.
In Sec.~(\ref{sec:RG_for_P_functions}) it is shown that
simply applying the RG transformation many times does not
identify functions that
can be written as the sum of a low-order polynomial plus a
contribution that is nonzero on a small fraction of the inputs.
One can identify functions of this type
by examining the set of functions whose outputs differ from the original
one on a small fraction of the input configurations ---
one of the functions in the set will be a low-order polynomial.
Sec.~\ref{sec:discussion} discusses the results in the framework
of phase transitions in condensed matter systems,
which renormalization group transformations
are typically used to study, and also discusses how the strategy
discussed here avoids the difficulties of ``natural proofs''
described in Ref.~\cite{razborov94natural}.
Sec.~\ref{sec:conclusions} presents the conclusions.
Appendix A presents the arguments demonstrating why
it is plausible most functions
that can be computed with polynomially
bounded resources can be written as a low-order polynomial
plus a term that is nonzero for a fraction of input
configurations that is exponentially small in $N/\log(N)$,
and discusses the non-generic nature of the
functions in P that do not have that property.
Appendix B shows that a typical Boolean function
cannot be written as a low-order polynomial plus a term
that is exponentially small in $N/\log(N)$.

\section{Renormalization group transformation}
\label{sec:RG}
The renormalization group (RG) procedure we define
takes a given function of $N$
variables and generates a function of $N-1$ variables~\cite{goldenfeld92,kadanoff66,wilson71,wilson79,white92}.
The variable that is eliminated is called the ``decimated" variable.
The procedure can be iterated, mapping a function of $N-1$ variables
into one of $N-2$ variables, etc.

The  transformation studied here
specifies whether the original
function's value changes if a given input variable
is changed.
Specifically, given a function $f(x_1,\ldots,x_N)\equiv f(\xvec)$, we define
\begin{eqnarray}
&~&g_{i_1}(x_1,x_1,\ldots,x_{i_1-1},x_{i_1+1},\ldots,x_N)
\equiv g_{i_1}(\xvec^\prime)\nonumber\\*
&~&~~=f(x_1,x_2,\ldots,x_{i_1-1},0,x_{i_1+1},\ldots,x_N)
\oplus
f(x_1,x_2,\ldots,x_{i_1-1},1,x_{i_1+1},\ldots,x_N)~,
\label{eq:g_1}
\end{eqnarray}
where $\oplus$ denotes addition modulo $2$~\cite{arithmetic_reference},
and the vector $\xvec^\prime$ denotes
the set of undecimated variables.
The function $g_{i_1}(x_1,x_2,\ldots,x_{i_1-1},x_{i_1+1},\ldots,x_N)$
is one if the output of the function $f$ changes when the value of the
decimated variable $x_{i_1}$ is changed
and zero if it does not.
Once $g_{i_1}$ has been obtained, the procedure can be repeated and
one can define $g_{i_1,i_2}$ as
\begin{eqnarray}
&~&g_{i_1,i_2}(x_1,x_1,\ldots,x_{i_1-1},x_{i_1+1},\ldots,x_{i_2-1},x_{i_2+1},x_N) 
\equiv g_{i_1,i_2}(\xvec^\prime)
\nonumber\\*
&&=~~g_{i_1}(x_1,x_2,\ldots,x_{i_2-1},0,x_{i_2+1},\ldots,x_N) \nonumber\\*
&&~~\oplus
g_{i_1}(x_1,x_2,\ldots,x_{i_2-1},1,x_{i_2+1},\ldots,x_N)\nonumber\\*
&&= ~~~f(x_1,x_2,\ldots,x_{i_1-1},0,x_{i_1+1},\ldots,x_{i_2-1},0,x_{i_2+1},\ldots,x_N)\nonumber\\*
&& ~~~\oplus
f(x_1,x_2,\ldots,x_{i_1-1},0,x_{i_1+1},\ldots,x_{i_2-1},1,x_{i_2+1},\ldots,x_N)
\nonumber\\*
&& ~~~\oplus f(x_1,x_2,\ldots,x_{i_1-1},1,x_{i_1+1},\ldots,x_{i_2-1},0,x_{i_2+1},\ldots,x_N)\nonumber\\*
 &&~~~\oplus
f(x_1,x_2,\ldots,x_{i_1-1},1,x_{i_1+1},\ldots,x_{i_2-1},1,x_{i_2+1},\ldots,x_N)~,
\end{eqnarray}
where the sums all denote addition modulo two.
The function $g_{x_{i_1},\ldots,x_{i_m}}(\xvec^\prime)$
obtained by decimating the $m$
variables $x_{i_1},\ldots,x_{i_m}$
does not depend on the order in which
the variables are decimated.

First we examine functions for which each of the coefficients
$A_{\alpha_1,\alpha_2,\ldots,\alpha_N}^{(0)}$ in Eq.~(\ref{eq:general_form})
is an independent random
variable chosen to be one with probability $p_0$ and zero with
probability $q_0=1-p_0$, where $0<p_0<1$.
We consider the sequence of functions obtained by successive
application of the renormalization group transformation
to such a generic random function.
The coefficients 
$A^{(i_1)}_{x_1,\ldots,x_{i_1-1},x_{i_1+1},\ldots,x_N}$
that characterize
the function $g_{i_1}(\xvec^\prime)$ obtained by decimating the variable
$i_1$ via Eq.~(\ref{eq:g_1}) are
\begin{eqnarray}
A^{(i_1)}_{x_1,\ldots,x_{i_1-1},x_{i_1+1},\ldots,x_N} 
= A_{x_1,\ldots,x_{i_1-1},0,x_{i_1+1},\ldots,x_N} \oplus
A_{x_1,\ldots,x_{i_1-1},1,x_{i_1+1},\ldots,x_N}~.
\end{eqnarray}
The original $A^{(0)}$'s are uncorrelated random variables, so it follows
that the $A^{({i_1})}$'s are independent random variables that are
one with probability $p_1=2p_0q_0$ and 
zero with probability $1-p_1$.
After $\ell$ iterations (after which $\ell$ variables
have been eliminated), the coefficients are still uncorrelated random variables,
and they are now one with probability $p_\ell$ and zero with probability $1-p_\ell$,
where the $p_\ell$ satisfy the recursion relation
\begin{eqnarray}
p_{\ell+1}=2p_\ell(1-p_\ell)~.
\label{eq:p_recursion}
\end{eqnarray}
The solution to Eq.~(\ref{eq:p_recursion}) is
\begin{eqnarray}
p_\ell = \frac{1}{2}\left ( 1 - (1-2p_0)^{2^\ell} \right ).
\label{eq:p_flow_result}
\end{eqnarray}
For any $p_0$ satisfying $0<p_0<1$, the values of the $p_\ell$
``flow'' as $\ell$ increases and eventually
approach
the ``fixed-point value'' of $1/2$~\cite{goldenfeld92}.
This behavior is exactly analogous to that displayed by
the partition functions describing thermodynamic phases in
statistical mechanical systems,
and so we interpret this behavior as evidence that there is
a phase of generic Boolean functions.

In Sec.~(\ref{sec:RG_for_P_functions}) we will be considering values of
$p_0$ that are very small but nonzero, for which case $p_\ell$
grows exponentially with $\ell$:
\begin{equation}
p_\ell = 2^\ell p_0\qquad({\rm when~}p_\ell \ll 1)~.
\label{eq:small_p_flow_result}
\end{equation}
After many renormalizations such functions will ``flow''
to the generic fixed point, so they are in the generic phase.
If one chooses $p_0=P(N)2^{-N}$, where $P(N)$ is a polynomial
in $N$, the function can be specified with polynomially bounded
resources by enumerating all input configurations for which
the function is nonzero.

Note that when the RG transformation is
applied to a generic Boolean function,
all the functions that are generated yield an output that is zero on a
fraction of the inputs that differs from $1/2$ by an amount that is
exponentially small in $N$. 
This follows because almost all Boolean functions have an
initial value of $p_0$ that differs from $1/2$ by an amount that is
the square root of the number of values chosen, or
$(2^{N})^{1/2}=2^{N/2}$.
Since all the $p_\ell$  deviate from $1/2$ by an amount
that is exponentially small in $N$, and
since the number of independent input configurations
remains exponentially large in $N$ until the number of
decimated variables is of order $N$,
for every function obtained via the renormalization transformation,
the fraction of input configurations yielding zero
deviates from $1/2$ by an amount that is exponentially small in $N$.

We next demonstrate that Boolean functions
that can be written as polynomials of degree of $\xi$ or less
when $\xi < N$ have the property that they yield zero after
$\xi+1$ renormalizations, for any choice of the decimated variables.

First we examine a simple example.
The parity function $\mathcal{P}(x_1,\ldots,x_N)$, which
is $1$ if an odd number of input variables are 1 and $0$ if
an even number of the 
input variables are 1~\cite{furst84,yao85,hastad86,wigderson06},
can be written as
\begin{equation}
\mathcal{P}(x_1,\ldots,x_N)=x_1\oplus x_2 \oplus \ldots \oplus x_N~.
\end{equation}
There are many less efficient ways to write the parity function,
but the result of the renormalization procedure does not depend on how
one has chosen to write the function, since it can be computed knowing
only
the values of the function for all different input configurations.
For the parity function, one finds, for any choice of decimated variables
$x_{j_1}$ and $x_{j_2}$, the functions resulting from one
and two renormalizations, $g^P_{j_1} (\xvec^\prime)$
and $g^P_{i_{j_1,j_2}}(\xvec^\prime)$, are:
\begin{eqnarray*}
&& g^P_{j_1} (\xvec^\prime) = x_{j_1} \oplus (1-x_{j_1}) = 1~;\\
&& g^P_{i_{j_1,j_2}}(\xvec^\prime) = 0~.
\end{eqnarray*}
Thus, applying the renormalization transformation to the parity function
yields zero after two iterations, in contrast to the behavior of a generic
Boolean function.

More generally, for any term of the form $T=y_{i_1} y_{i_2} \ldots y_{i_m}$,
with $y_i=x_i$ or $1-x_i$, the quantity $T(x_i=1) \oplus T(x_i=0)$ is either
zero (if $y_i$ does not occur in $T$) or else is the
product of $m-1$ instead of $m$ of the $y$'s; for example
\begin{equation}
T(y_{i_1}=1) \oplus T(y_{i_1}=0)=y_{i_2} \ldots y_{i_m}~.
\end{equation}
Because the effect of the RG procedure on the sum of terms is equal
to the sum of the results of the transformation applied to the
individual terms,
any function that is the mod-2 sum of terms that are all
products of fewer than $m$ $y$'s will yield zero after $m$ renormalizations,
for any choice of the decimated variables.
It follows immediately that a function that is a polynomial of degree
$\xi$ or less has the property that applying the RG transformation to
it $\xi+1$ times yields zero for any choice of the decimated variables.

This result demonstrates that the RG transformation
distinguishes generic Boolean functions from functions that can be
written as polynomials of degree $\xi$ or less, when $\xi<N$.
The qualitatively different behavior upon renormalization of polynomials
of degree $\xi$ from generic Boolean functions can be interpreted as
evidence that these two classes of functions are in different phases.

We now demonstrate that the RG method also identifies as non-generic
functions that depend on a
composite quantity such as the arithmetic sum of the variables.
Functions in P with this property  include
MAJORITY (which is one if more than half the inputs are
set to one, and zero otherwise)~\cite{razborov87}
and DIVISIBILITY MOD p (which is one if the number of
inputs that are set to one is divisible by an odd prime p and
zero otherwise)~\cite{smolensky87,smolensky93}.
The renormalization group approach distinguishes
such functions from generic Boolean functions because
the output of all the functions in the sequence
is constrained to be identical for very large
sets of input configurations.
We first show that MAJORITY and DIVISIBILITY MOD p are
both distinguished from a generic Boolean function by the renormalization
group procedure, and then we argue that the RG
procedure distinguishes any function of the
arithmetic sum of the inputs from a generic Boolean
function.
We expect that the argument will be
generalizable to apply to a broad class of functions that depend on
other composite quantities that are
specific combinations of the input variables.

First we consider the behavior when the RG transformation
is applied to DIVISIBILITY MOD 3.
Since this function is nonzero when the arithmetic sum
$\sum_{j=1}^N x_j$ is divisible by $3$, changing an input $x_i$
changes the output value
when the sum of the
other input variables is either zero or two.
Thus, the renormalized function
$g_i(\vec{x}^\prime)$ is nonzero for any $i$ on
a fraction of the input
configurations that is very close to $2/3$.
Every succeeding renormalization also yields
a function that is nonzero when
the sum of the remaining variables is either zero or two.
This behavior differs from that of a generic
Boolean function, in which the renormalized functions are nonzero
for a fraction of inputs that is very close to $1/2$.
More generally, when the RG is applied to DIVISIBILITY MOD p, with
p an odd prime, the behavior of the sequence of functions is
determined by the value of the mod p remainder of the undecimated
variables.
The functions in the sequence yield the output one
when the remainder mod p takes on certain values, and typically,
after a small number of iterations, these
values cycle with a finite period.
Therefore, the fraction of input configurations that lead to a nonzero
input essentially cycles also (the cycling is not exact only because the
fraction of input configurations with a given value of the remainder mod
p changes very slightly with $N$), and, since p is odd, none of the
fractions in the cycle is close to $1/2$.

The behavior obtained when the RG procedure is applied to the MAJORITY
function is also significantly different from that of a generic Boolean function.
The first renormalization step yields a function that is nonzero when
the sum of the undecimated variables is $N/2-1$, and the second step
yields a function that is nonzero when the sum of the undecimated variables
is either $N/2-2$ or $N/2-1$.
The functions obtained after $j$ decimations are nonzero on a fraction of
inputs that is bounded above by $C j/\sqrt{N}$, where $C$ is a constant
of order unity, so long as $j\ll\sqrt{N}$.
The original function is thus identified as non-generic because so long as
the number of renormalizations applied is much smaller than $\sqrt{N}$
the renormalized functions are all nonzero
on a fraction of input configurations that is much less than $1/2$.

Next we argue that the renormalization group approach distinguishes
any function of the arithmetic sum of the inputs from a generic Boolean function.
The physical intuition underlying the argument is that all the functions
in the sequence depend only on the arithmetic sum of the undecimated
variables, and when the number of undecimated variables is
$\mathcal{N}$, the number of configurations of the undecimated variables
whose arithmetic sum is constrained to be $\mathcal{S}$, is
$\mathcal{N}!/\mathcal{S}!(\mathcal{N}-\mathcal{S})!$.
One can use Stirling's series~\cite{marsaglia90}
to show explicitly that when $N$ is large,
then the number of configurations with a given value of $\mathcal{S}$
is a polynomial in $1/N$ times $2^N$ for a number of values of $\mathcal{S}$
that grows as the square root of $N$.
Therefore, the {\em differences} in the fraction of configurations
yielding different values of $N$ decay polynomially with $N$, and
the fraction of input configurations yielding one should either
be exactly 1/2 or else must deviate from 1/2 by an amount that decreases
only polynomially with $N$.

\section{Renormalization procedure for characterizing functions
that can be constructed using polynomially bounded resources.}
\label{sec:RG_for_P_functions}
\label{sec:RG_for_low_order_polynomial_functions}
This section addresses the relationship between
non-generic phases of Boolean functions and the computational
complexity class P of functions that can be computed with polynomially
bounded resources.

There are functions that are in P that are neither
polynomials of degree $\xi$ with $\xi<N$ nor functions of composite
variables.
For example, because the sum of two functions that are in P is in P,
a sum of any function that is in P with
a small ``generic'' piece specified by Eq.~(\ref{eq:general_form}) with the
coefficients chosen independently and randomly to be one with probability
$p_0=\mathcal{P}(N)2^{-N}$, where $\mathcal{P}(N)$ is a polynomial
in $N$, is in P.
Eq.~(\ref{eq:small_p_flow_result}) shows that $\ell$
renormalizations
cause the value of $p_\ell$ to grow exponentially with $\ell$, $p_\ell=2^\ell p_0$;
in renormalization group parlance~\cite{goldenfeld92}
the remainder is
a ``relevant'' perturbation.
Since the generic piece renormalizes towards the generic fixed point
at which exponentially close to half the inputs yield a nonzero output,
whether or not the function resulting from many renormalizations
can be identified as non-generic depends on whether the first piece
yields a nonzero result after many renormalizations.
A function of a composite variable yields a result different both from zero and
from that of generic functions, and when a small generic piece 
is added to such a function, renormalization still yields a non-generic result.
However, because after $\xi+1$ renormalizations of a
polynomial of order $\xi$ one obtains zero,
renormalizing functions that are the sum of a low-order polynomial and a small
generic piece yields zero plus the generic result, and
so cannot be identified as non-generic by straightforward application
of the renormalization transformation.

The number of polynomials of $N$ variables with degree $\xi$ is
$2^{\sum_{k=1}^\xi N!/(\xi!(N-\xi)!)}$~\cite{polynomial_count},
which when $\xi\ll N$ can be approximated as
$2^{e(N/\xi)^\xi}$.
Therefore, when $\xi$ scales as a fractional
power of $N$,
there are many
more polynomials of degree $\xi$ than there are functions in P.
On the other hand, the product of all $N$ variables
$x_1\ldots x_N$ is in P, so there are functions in P that cannot be
written as polynomials of degree $\xi$ for any $\xi<N$.
Therefore, using our definition of a phase based on the
behavior yielded by repeated renormalization, P is not
a phase.
There are non-generic functions that are not in P and there are functions
in P that are in the generic phase.
However, note that a product of $M$ variables
is nonzero for only a fraction $2^{-M}$ of the
input configurations.
For example, the term $x_1x_2\ldots x_R$ is nonzero only for
input configurations that have
$x_1=x_2=\ldots=x_R=1$.
The sum of a polynomially large number $M$ of terms of this type
is nonzero only on a fraction of inputs that is bounded above
by $M/2^R$.
In Appendix A it is argued that the
functions in P that are in the generic phase have the property 
that for any $\xi<N$,
any Boolean function of $N$ variables $f(x_1,\ldots,x_N)$ that
is in P can be written as the sum:
\begin{equation}
f(x_1,\ldots,x_N)=\mathcal{P}_\xi(x_1,\ldots,x_N)\oplus \mathcal{R}_\xi(x_1,\ldots,x_N)~,
\label{eq:bound_for_fns_in_P}
\label{eq:decomposition}
\end{equation}
where $\mathcal{P}_\xi(x_1,\ldots,x_N)$ is a polynomial of
degree no more than $\xi$ and
the remainder term $\mathcal{R}_\xi(x_1,\ldots,x_N)$ is nonzero on a fraction
of input configurations that is bounded above by
$\mathcal{C}2^{-\alpha\xi/\log_2(N)}$,
with $\mathcal{C}$ and $\alpha$ positive constants.

As discussed above,
using the RG transformation to identify functions that
satisfy Eq.~(\ref{eq:decomposition})
is not entirely straightforward --- the obvious strategy,
seeing if the functions obtained after renormalizing $\xi+1$ times 
have a small remainder term, fails
because renormalization yields exponential growth in
the fraction of input configurations for
which the remainder term is nonzero.
This difficulty can be circumvented
by examining {\em all} functions that differ
from the function in question
on a fraction of
input configurations no greater than $\mathcal{C}2^{-\alpha \xi/\log_2(N)}$.
If the original function obeys Eq.~(\ref{eq:decomposition}), then
one of the ``perturbed'' functions will have a remainder
term that is zero, and applying
the renormalization transformation to it $\xi+1$ times yields zero
for all choices of the decimated variables.

There are functions known to be in P that
can written as the sum of a function of a composite variable
plus a function that is nonzero on a small fraction of inputs.
Nongeneric behavior is obtained upon renormalization for
all such functions except for those for which
all functions in the renormalization sequence
yield one for exactly half the input configurations.
The procedure for identifying such functions is exactly analogous 
as for identifying functions that can be approximated as low-order polynomials ---
examine the properties under renormalization
of all the functions that are yield the same output as the one in question except
for a small fraction of the inputs.

%
Finally, we note that in Appendix B it is demonstrated that
almost all generic random functions do not satisfy
Eq.~(\ref{eq:decomposition}) when $\xi$ scales as a fractional
power of $N$.

\section{Discussion}
\label{sec:discussion}
This paper presents a renormalization group approach that distinguishes
generic Boolean functions of $N$ variables from functions that
can be written as a polynomial
of degree $\xi$, with $\xi \ll N$,
and also from functions that depend only on composite quantities
such as the arithmetic sum of all the input variables.
The method provides a consistent framework
for identifying many
different functions as non-generic.

The renormalization group approach 
also provides a natural framework for understanding
why the P versus NP question is so difficult.
Functions computable
with polynomial resources do not comprise a phase --- there are
functions that are in a non-generic phase that are not in P, and
there are functions in P for which
the renormalization group yields a
``flow'' that is towards the generic fixed
point and hence are in the ``generic'' phase.
The possibility of using the RG approach to demonstrate that a given
Boolean function is not in P
arises because it is possible that all functions in P that are in the generic
phase are all close to a phase boundary of a non-generic phase.
Whether the renormalization group approach can provide a means
for determining whether or P is distinct from NP depends on
whether it is possible to demonstrate that all efficiently computable
functions are in or near a non-generic phase.

\begin{figure}[htbp]
\begin{center}
\includegraphics[angle=90,height=5cm]{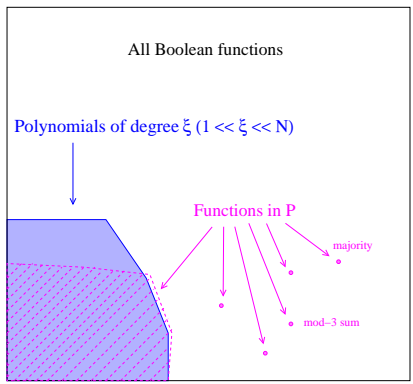}
\caption{Schematic phase diagram for Boolean functions.
Within the set of all Boolean functions of $N$ Boolean variables
there is a generic phase,
a phase
consisting of functions that can be written as polynomials of order no
greater than $\xi$
with $\xi \ll N$, and
there are phases corresponding to functions of composite
variables such as the arithmetic sum of all the inputs.
Some polynomials of degree $\xi$ are not in P, and
some functions that can be computed with
polynomial resources cannot be written either
as polynomials of degree
$\xi$ for any $\xi<N$ or as functions of a composite
variable.  Therefore, P does not denote a phase.
However, we conjecture that that all functions in P
are either in a non-generic phase or else
very close to the low-order-polynomial phase boundary.
}
\label{fig:phase_diagram}
\end{center}
\end{figure}

The procedure used here of using the behavior yielded by a
renormalization group
transformation to identify different phases of Boolean functions
is entirely analogous to
a procedure
presented by Wilson~\cite{wilson79} to identify different
thermodynamic phases of the Ising model, used to describe
magnetism in solids.
Wilson showed that individual
configurations of Ising models could be identified as being in
either a ferromagnetic phase or paramagnetic phase by repeatedly
eliminating spins
and examining the resulting configurations --- if after many
renormalizations all the
spins are aligned, then the system is in the ferromagnetic phase,
while if after many renormalizations the spin orientations
are random, then the system is in the paramagnetic phase.
Viewing the analogy between the results for magnets
and
the qualitatively different behavior of the renormalization
group
``flows'' for polynomials of degree $\xi$, for functions of
composite variables, and for generic Boolean
functions as an indication that low-degree polynomials
and functions of composite variables are both
non-generic ``phases," we propose
the schematic phase diagram for
Boolean functions,
shown in Fig.~\ref{fig:phase_diagram}.

If it can be shown that all functions in P are either in a
non-generic phase or else very close to a phase boundary,
then the procedure described here leads to
a specific algorithmic approach to the
P versus NP question --- if a given function that is
obtained as the answer to a problem in NP fails to
be close enough to a non-generic phase, then
one has shown that P is not equal to NP.
(Ref.~\cite{coppersmith06b} 
advocates a family of candidate
functions for testing using the strategy proposed in
this paper, but the strategy can be implemented
for any candidate function.)
Appendix B shows that almost all Boolean functions are
not close to non-generic phase boundaries.
Appendix A argues that
the construction of a function
in P that does not satisfy Eq.~(\ref{eq:decomposition})
requires delicate balancing that may signal
the existence of a composite variable, but the argument
is only speculative.
Progress on this issue is the key
to using the RG approach to be able to address the
P versus NP question.

Because the procedure discussed in 
Sec.~\ref{sec:RG_for_P_functions} 
requires a number of operations that scales superexponentially
with N, the procedure proposed here is not
a ``natural proof" as
discussed in Ref.~\cite{razborov94natural} and therefore
does not yield a method for breaking
pseudorandom number generators.
However,
direct numerical implementation of the procedure is not likely
to be
computationally feasible.

\section{Conclusions}
\label{sec:conclusions}
This paper presents a renormalization group approach that can be
used to distinguish a generic Boolean function from
(1) a Boolean function of $N$ variables
that can be written
as a polynomial of degree $\xi$ with $\xi<N$, and
(2) a function that depends only on a composite variable
(such as the arithmetic sum of the inputs).
An algorithm for determining whether a function
differs from a polynomial of degree $\xi$ on a fraction
of inputs that is exponentially small in $\xi/\log(N)$ is presented.
The possible relevance of these results to the question of whether
P and NP are distinct is discussed.

\section{Acknowledgments}
\label{sec:acknowledgments}
The author is grateful to Prof. Daniel Spielman for pointing out a serious
error in the original version of the manuscript, and for
support from NSF grants CCF 0523680 and DMR 0209630.

\vspace{.3cm}
\noindent{\bf \Large Appendix A~~~
Characterization of the functions that can be constructed with
a polynomially large number of operations.
}

In this appendix we examine the properties
of functions that can be computed with polynomially
bounded resources.
First we discuss why it is plausible that almost all functions
in P can
written in the form Eq.~(\ref{eq:decomposition}),
which is the sum of two terms,
the first a polynomial of degree $\xi$,
and the second a
correction term that is nonzero on a
fraction of input configurations that is exponentially small in $\xi/\log(N)$.
We then examine
known functions in P that cannot be written in this form,
arguing that they have special properties that may give rise to
the emergence of a composite variable on which the function
depends, which would lead to non-generic behavior 
upon renormalization. 

To see why it is hard to construct functions in P that do not satisfy
Eq.~(\ref{eq:decomposition}),
we consider the process by which functions can be constructed.
First we show that a starting polynomial that is the sum of
polynomially many terms whose factors are all either $x_i$ or $(1-x_i)$ satisfies
Eq.~(\ref{eq:decomposition}).
Then we show that the sum of two functions that each 
obey Eq.~(\ref{eq:decomposition}) also satisfies Eq.~(\ref{eq:decomposition}),
and also that the coefficient multiplying the correction term grows
sufficiently slowly that the bound remains true even after
a number of additions that grows polynomially with $N$.
We then consider products of such functions.
The behavior is more complicated, but we argue that
a similar decomposition works in most circumstances
because when many terms are multiplied together,
the result is nonzero only
on a small fraction of inputs.
Finally, we examine some functions in P which do
not satisfy Eq.~(\ref{eq:decomposition}) and note that
they involve a delicate balance that enables the sum of
a finite number of products to be nonzero on the  same
fraction of inputs as the individual terms.
It is plausible that this nongeneric property is associated with
the nongeneric behavior of these functions upon renormalization.

First consider a polynomial 
$A(x_1,\ldots,x_N)$
that is the mod-2 sum of polynomially many
terms that are all of the form $y_{i_1}\ldots y_{i_m}$,
where $y_i$ is either $x_i$ or $1-x_i$:
\begin{eqnarray}
{A}(x_1,\ldots,x_N)=C_0 + \sum_{\eta=1}^N
\sum_{k_\eta=1}^{M_\eta} 
y_{i_1(\eta ,k_\eta)} \ldots y_{i_{\eta}(\eta, k_\eta)}~.
\end{eqnarray}
Here, $C_0$ is a constant,
$\eta$ denotes the number of factors of $y_i$ in a term, $k_\eta$
is the index labeling the different terms with $\eta$ factors, 
$i_j(\eta,k_\eta)$ denotes the index of the $j^{th}$ factor
in the term $k_\eta$, and
each $M_\eta$, the number of terms with $\eta$ factors,
is bounded above
by a polynomial of $N$.
We will obtain
bounds on the number of
configurations for which the output is nonzero by
considering standard addition instead of
modulo-two addition,
which means that we will
overcounting by
including configurations for which an even number
of terms in the polynomial expansion are nonzero. 
Each term with
$\eta$ factors is nonzero only on a
fraction $2^{-\eta}$ of the inputs.
Therefore, if we define $\rho_{A}(\eta)$ to be
the fraction of inputs of
$A(x_1,\ldots,x_N)$
for which the sum of all the terms with
$\eta$ factors is nonzero, we have
\begin{equation}
\rho_A(\eta) \le C_A 2^{-\alpha \eta}~,
\label{eq:decay_condition}
\end{equation}
for constant $C_A$ and $\alpha=\frac{1}{2}-\epsilon$,
with $\epsilon$ infinitesimal.

Now consider the addition of two functions
$P(x_1,\ldots,x_N)$ and
$Q(x_1,\ldots,x_N)$ that satisfy
Eq.~(\ref{eq:decomposition}) for positive $\mathcal{C_P}$,
$\mathcal{C_Q}$, and $\alpha$.
Again we consider standard addition instead of
modulo-two addition.
Because the sum
$S(x_1,\ldots,x_N)=P(x_1,\ldots,x_N)+Q(x_1,\ldots,x_N)$
has the property that all terms in the sum appears
in at least one of the summands, we have
\begin{equation}
\rho_S(\eta) \le \rho_P(\eta)+\rho_Q(\eta)~;
\end{equation}
the sum obeys Eq.~(\ref{eq:decay_condition}) with
the same value of $\alpha$ and with $C_S \le C_P+C_Q$.
Adding polynomially many terms can increase the prefactor
only by an amount that grows no faster than polynomially in $N$.

We next consider the product of two functions that satisfy
Eq.~(\ref{eq:decay_condition}).
We write
\begin{eqnarray}
A(\vec{x}) &=& P_A^\xi(\vec{x})+R_A^\xi(\vec{x})\nonumber\\
B(\vec{x}) &=& P_B^\xi(\vec{x})+R_B^\xi(\vec{x})~,
\end{eqnarray}
where $P_A^\xi$ and $P_B^\xi$ are polynomials of order $\xi$
with $T_A$ and $T_B$ terms respectively, and
$R_A^\xi(\vec{x})$ and $R_B^\xi(\vec{x})$ are both nonzero
on a fraction of inputs that is less than $\mathcal{C} 2^{-\alpha \xi}$ for
positive constants $\mathcal{C}$ and $\alpha$.

We write the product of $A(\vec{x})$ and $B(\vec{x})$ as
\begin{eqnarray}
D(\xvec) &=& A(\xvec)B(\xvec)\nonumber\\
&=& (P_A^\xi(\xvec)+R_A^\xi(\xvec))(P_B^\xi(\xvec)+R_B^\xi(\xvec))\nonumber\\
&=& P_A^\xi(\xvec)P_B^\xi(\xvec)+P_A^\xi(\xvec)R_B^\xi(\xvec)
+R_A^\xi(\xvec)P_B^\xi(\xvec)+R_A^\xi(\xvec)R_B^\xi(\xvec)~.
\end{eqnarray}
Now $P_A^\xi(\xvec)R_B^\xi(\xvec)$ is nonzero on fewer inputs than
$R_B^\xi(\xvec)$ (this follows since a product is nonzero only if
each of its factors is nonzero), and, similarly, $R_A^\xi(\xvec)P_B^\xi(\xvec)$
is nonzero on fewer inputs than either $R_A^\xi(\xvec)$
or $R_A^\xi(\xvec)$, so the sum of the
last three terms
must be less
than $3\mathcal{C} 2^{-\alpha\xi}$.
Therefore, these contributions to the remainder term in the
product remain exponentially
small, with a coefficient that remains bounded by a polynomial in $N$
after polynomially many multiplications.
Therefore, it only remains to consider the properties of the product
$P_A^\xi(\xvec)P_B^\xi(\xvec)$, which we write
\begin{equation}
P_A^\xi(\xvec)P_B^\xi(\xvec)=P_D^\xi(\xvec)+R_D^\xi(\xvec)~,
\label{eq:P_Dequation}
\end{equation}
where $P_D^\xi(\xvec)$ is a polynomial of degree $\xi$
and $R_D^\xi(\xvec)$ is a remainder term that we need to bound.

To bound the magnitude of the remainder,
let us multiply out the polynomials in Eq.~(\ref{eq:P_Dequation}) so that
they are all sums of terms that are
products of the form $y_{i_1}\ldots y_{i_j}$, terms that
we will denote as  ``primitive."
Let $T_A$ be the number of primitive terms in $P_A^\xi(\xvec)$, and
$T_B$ be the number of primitive terms in $P_B^\xi(\xvec)$.
Note that every primitive term in the product with more than $\xi$ factors
is nonzero on a fraction $2^{-\xi}$ or less of the input configurations.

Since the total number of primitive terms in $R_D^\xi(\xvec)$
is bounded above by
$T_AT_B$, the fraction of inputs on which the sum of the
terms with at least $\xi$ factors
is nonzero is bounded above by $T_AT_B2^{-\xi}$.
So long as $T_A$ and $T_B$ are both less than exponentially
large in $\xi$, then 
this remainder term is exponentially small in $\xi$.
The multiplication process must start with values
of $T_A$ and $T_B$ that are both bounded by a polynomial
of $N$, but because multiplications can be composed,
we need to examine the behavior of
$T_D$, the number of primitive terms
in $P_D^\xi(\xvec)$.

A simple upper bound for $T_D$ is obtained by ignoring
all possible simplifications that could reduce
the total number of terms in the product:
\begin{equation}
T_D \le T_AT_B~.
\end{equation}
This equation describes geometric growth.
If $\mathcal{M}$ polynomials are multiplied together, all of which
have fewer than $CN^Y$ terms
for fixed $C$ and $Y$,
then the total number of terms in the product,
$T_\mathcal{M}$,
satisfies the bound 
\begin{equation}
T_\mathcal{M} \le  (CN^{Y})^ \mathcal{M}~.
\label{eq:largeMbound}
\end{equation}
This bound on the number of terms in the
product is
much smaller than $2^\xi$ so long as $\mathcal{M}$
satisfies
\begin{equation}
\mathcal{M} \ll \xi/ (Y\log_2 N+\log_2 C)~.
\end{equation}

A useful bound on multiplicative terms that are products of more
than $\xi/(Y\log_2 N) $ factors can be obtained by exploiting the fact that
the product of two functions is nonzero for
a given input only if each of the factors is.
Specifically, consider the product $AB$, and say that $A$ is nonzero
on a set of $M_A$ inputs.
If $B$ is nonzero on less than a fraction $\sigma$
of the inputs in this set for some $1/2<\sigma<1$, then the
product $AB$ is nonzero on fewer than $\sigma M_A$ inputs,
and if not, then the product $A(1-B)$ is nonzero
on fewer than $(1-\sigma)M_A$ inputs,
and one can write
$AB=A+A(1-B)$.~\cite{proliferation_footnote}

The result of $\mathcal{M}$ multiplications is then nonzero only
on a fraction of inputs bounded above by $2^{-\mathcal{M}\log_2\sigma}$.
Therefore, a product of more than $\xi/Y\log_2(N)$ factors
is nonzero on no more than a fraction $2^{-\tilde{C}\xi/\log_2(N)}$
of the inputs, where $\tilde{C}$ is a positive constant,
and the entire product can be moved into the
remainder term.

The arguments above indicate that the remainder term tends to
be small for products
because the number of terms in the
polynomial that are of order $\xi$ or less can be bounded for
products of small numbers of terms, and products of many terms
are nonzero on a small enough fraction of the input configurations
that they can be considered to be part of the remainder term.
However, there are functions in P that do not obey
Eq.~(\ref{eq:decomposition}).
Two examples of functions that are in P that have been proven
to violate Eq.~(\ref{eq:decomposition}) are MAJORITY (which is one
when more than half input variables have been set to one
and zero otherwise)~\cite{razborov87}
and DIVISIBILITY MOD p, which is one if the
sum of the input variables is divisible by
an odd prime p~\cite{smolensky87,smolensky93}.
Both these functions depend only on the arithmetic (not mod-2) sum of
all the variables, $x_1+x_2+\ldots x_N$.
Calculating the sum of $N$ variables can be done with polynomially
bounded resources because one
need only keep track of a running sum, which is the same for
many different values of the individual $x_j$.
For instance, when $k=N$, 
there are $N!/((N/2)!)^2 \approx 2^N/\sqrt{2\pi N}$ different
ways to choose the $x_1\ldots x_k$ so that their sum is $N/2$.

It is instructive to consider an algorithm for computing DIVISIBILITY MOD $3$
to see how the function avoids being a low order polynomial.
Some pseudocode for a simple algorithm for this problem is:
\begin{eqnarray*}
&& {\rm divisibility~ mod~ 3:}\\
&&~~~~~	{\rm start: remainder0[0]=1, remainder1[0]=remainder2[0]=0}\\
&&~~~~~	{\rm for~ each~ i>0}\\
&&~~~~~	{\rm remainder0[i+1] = remainder0[i]*(1-x_{i+1})\oplus remainder2[i]*x_{i+1}}\\
&&~~~~~	{\rm remainder1[i+1] = remainder1[i]*(1-x_{i+1})\oplus remainder0[i]*x_{i+1}}\\
&&~~~~~	{\rm remainder2[i+1] = remainder2[i]*(1-x_{i+1})\oplus remainder1[i]*x_{i+1}}\\
&&~~~~~	{\rm answer=remainder0[N]}
\end{eqnarray*}
The quantity remainder0[i]+remainder1[i]+remainder2[i]
is unity for every i, and the fraction of inputs for which each remainder
variable is nonzero is very close to $1/3$ and does not decay exponentially
with i.
The fractions do not decay or grow because the equation for
each remainder for a given $i$ is the sum of
two products.
The product ${\rm remainder0[i](1-x_{i+1})}$ is nonzero on
half the inputs on which remainder0[i] is nonzero, and similarly
for the other term ${\rm remainder2[i]*x_{i+1}}$.
Because ${\rm remainder0[i+1]}$ is the sum of two terms, each of which is
nonzero on almost exactly half the outputs for which
${\rm remainder0[i]}$ is nonzero, ${\rm remainder0[j]}$ remains of
order of but less than unity for all j.
It is plausible that
this exquisite cancellation leads to the existence of a composite variable
on which the function depends,
or, more generally, to non-generic behavior upon renormalization.
Because obtaining a function that cannot be written as a low-order polynomial
plus a term that is nonzero except for a small fraction of input configurations
requires a series of delicate cancellations, it is also
extremely plausible that the fraction of functions that are
in P and do not satisfy Eq.~(\ref{eq:decomposition}) is extremely small.

As discussed in the main text,
the renormalization group distinguishes functions that depend
on the sum of the values of the input variables from
generic Boolean functions because the renormalization transformation
preserves the property that
a given value for the composite variable occurs for an
exponentially large numbers of input configurations.
Moreover, at least when the composite variable is the
arithmetic sum of the inputs, the fractions of input configurations
for which the sum of the variables takes on different values differ by
an amount that decays only polynomially with $N$.
Therefore, such functions can yield one either on exactly half the
inputs or else on a fraction of the inputs that differs from $1/2$ by
an amount that is at least as large of $N^{-x}$ for some positive $x$.

To summarize, in this appendix we discuss the restrictions on Boolean
functions of $N$ variables
that can be computed with resources that are bounded
above by a polynomial in $N$.
Many functions in P have the property that they
can be written, for any fixed $\xi$,
as the sum of a polynomial of degree $\xi$ and a term that is
bounded above by $\mathcal{C}2^{-\alpha\xi/\log_2(N)}$ for positive
constants $\mathcal{C}$ and $\alpha$.
Known functions in P that cannot be approximated by low-order
polynomials have the property that they have a dependence on a
composite variable.
The renormalization group transformation provides a means for
distinguishing
both types of functions from generic Boolean
functions.

\vspace{.3cm}
\noindent{\bf \Large Appendix B:  Demonstration that a typical Boolean
function does not satisfy Eq.~(\ref{eq:bound_for_fns_in_P}).}

In this appendix it is shown that for a typical Boolean function,
changing the outputs for an exponentially small fraction of the
inputs does not yield a low-order polynomial.
Specifically, given a value of $\xi$ with $\xi\propto N^y$
with $0<y<1$,
if one changes the output value of a typical Boolean function
for no more than $\mathcal{C} 2^{N-\alpha\xi/\log_2(N)}$
input configurations, then the resulting function cannot be
written as a polynomial of degree $\xi$ or less.
This is done by showing
that the number of Boolean functions that
satisfy Eq.~(\ref{eq:bound_for_fns_in_P}) is
much less than the number of Boolean
functions of $N$ variables.

The number of Boolean functions of $N$ variables satisfying
Eq.~(\ref{eq:bound_for_fns_in_P}), $\mathcal{B}(N,\xi)$,
satisfies
\begin{equation}
\mathcal{B}(N,\xi) \le \mathcal{F}(N,\xi) \mathcal{M}(N,\xi)~,
\end{equation}
where
$\mathcal{F}(N,\xi)$ denotes
the number ways to choose up to
$\mathcal{C}2^{N-\alpha\xi/\log_2(N)}$ input configurations and
$\mathcal{M}(N,\xi)$ is the number of polynomials
of degree $\xi$.

Let $\Phi=\mathcal{C} 2^{N-\alpha\xi/\log_2(N)}$ be the maximum
number of configurations whose outputs we are allowed
to alter,
and $\Omega=2^N$ be the total
number of input configurations.
The quantity $\mathcal{F}(N,\xi)$ is the number of ways that one
can choose up to $\Phi$ items
out of $\Omega$ possibilities.
We have
\begin{eqnarray}
\mathcal{F}(N,\xi) &=&\sum_{s=1}^\Phi
 \frac{\Omega!}{s !(\Omega-s)!}\nonumber\\
&\sim& 
e(\Omega/s)^s
= e{(2^{\alpha \xi/\log_2(N)}/\mathcal{C} )}^{\mathcal{C}2^{N-\alpha\xi/\log_2(N)}}~,
\end{eqnarray}
where the last line applies when $1 \ll \xi \ll N$.
Next note that $\mathcal{M}_\xi$, the number of different polynomials
of degree less than or equal to $\xi$, is:
\begin{eqnarray}
\mathcal{M}_\xi &=& 2^{\sum_{j=0}^\xi N!/j!(N-j)!}\nonumber\\
&\sim& 2^{e(N/\xi)^\xi}~,
\label{eq:num_polynomials}
\end{eqnarray}
where again the last line assumes $1 \ll \xi \ll N$.
Eq.~(\ref{eq:num_polynomials}) follows 
because all polynomials of degree $\xi$ or less can be written
as a sum over all terms that are
products of the form $x_{i_1}\ldots x_{i_j}$
with $j\le\xi$.
There are
$\sum_{j=1}^\xi N!/[j!(N-j)!]$
such terms,
and each coefficient can be either $1$ or $0$.
Thus, when $1 \ll \xi \ll N$,
the total number of functions that satisfy
Eq.~(\ref{eq:bound_for_fns_in_P})
is bounded above by
\begin{eqnarray}
\mathcal{B}(N,\xi) &\le& \left ( 
e(2^{\alpha \xi/\log_2(N)}/\mathcal{C})^{\mathcal{C}2^{N-\alpha\xi/\log_2(N)}}
\right )
\left (2^{e(N/\xi)^\xi}\right )~,
\end{eqnarray}
which, as $N\rightarrow\infty$ and
$\xi \propto N^a$ with $0<a<$1, is much smaller than $2^{2^N}$,
the total number of Boolean functions of $N$ Boolean variables.

A second non-rigorous but informative
argument to see that generic Boolean functions
do not satisfy Eq.~(\ref{eq:bound_for_fns_in_P})
is to consider a generic Boolean function in which each coefficient
$A_{i_1,\ldots,i_N}$ is chosen independently and randomly
to be $1$ or $0$ with equal probability.
For a typical Boolean function, one can always find a configuration
satisfying Eq.~(\ref{eq:bound_for_fns_in_P}) by changing just
about half the output values so that the function has the same value
for all inputs.
The question is whether one can obtain $g_{x_{i_1},\ldots,x_{i_M}}(\xvec^\prime)=0$
for all choices of the $M$ decimated variables by changing the
function for many fewer configurations than that.
For a given $g$ in which $M$ variables have been decimated,
one can find a configuration satisfying
$g_{x_{i_1},\ldots,x_{i_M}}(\xvec^\prime)=0$
for the $2^{N-M}$ different possible ${\xvec^\prime}$
by changing the output for just about $2^{N-M-1}$ different
input configurations.
But one must arrange for $g_{x_{j_1},\ldots,x_{j_M}}(\xvec^\prime)$
to vanish
for all possible choices of the $M$ variables to be decimated.
There are $N!/[M!(N-M)!] \sim e(N/M)^M$ different ways to
choose the decimated variables, so a naive estimate is that
one must adjust $2^{N-M}$ configurations for each of
$e(N/M)^M$ choices of the decimated variables, or
$2^{N-M+1+M\log(N/M)}$, which exceeds $2^N$ for all 
$M \ll N$.
This argument is useful because it makes it clear why one must
examine all choices of the decimated variables to distinguish
functions that do not satisfy Eq.~(\ref{eq:bound_for_fns_in_P}).


\begin{thebibliography}{30}
\bibitem{polynomial_count}
To obtain the number of polynomials of degree $\xi$ or less,
note that each can be written as a sum of
terms of the form $x_{i_1}\ldots x_{i_k}$ for all $k \le \xi$.
There are $N!/k!(N-k)!$ ways to choose $k$ indices out of $N$
possibilities, so there are $\sum_{k=1}^\xi N!/k!(N-k)!$ different
possible terms in the polynomial, each of which occurs with a
coefficient of either one or zero.
Thus, there are $2^{\sum_{k=1}^\xi N!/k!(N-k)!}$ different
polynomials of degree $\xi$ or less.
\bibitem{proliferation_footnote}
	One might worry that
	products of the form $A_1A_2\ldots A_M$,
	where each $A_i$ is nonzero on more than half of the inputs, and
	$M$ is of order $N$, might pose a problem, for if one writes
	$A_1A_2\ldots A_M=(1-A_1^\prime)(1-A_2^\prime)\ldots(1-A_M^\prime)$,
	then the number of terms with $m$ factors is $M!/(M-m)!m!$, which
	can be as large as $2^{M/2}$ (when $m=M/2$).
	This term proliferation is not a problem if one chooses $\sigma$ to be
	strictly greater than $1/2$ (say, $3/4$), since the number of terms
	with a given number of terms in the product is overwhelmed by the
	decrease in the fraction of inputs for which each individual term is nonzero.
\end{thebibliography}
 \end{document}